\documentclass[reprint,preprintnumbers,nofootinbib,amsmath,amssymb,aps]{revtex4-1}
\usepackage{slashed}
\usepackage{graphicx}
\usepackage{dcolumn}
\usepackage{bm}

\usepackage{amsfonts}
\usepackage{latexsym}
\usepackage{dsfont}
\usepackage{amsmath}
\usepackage{amssymb}
\usepackage{amssymb}
\usepackage{amsthm}
\usepackage{xcolor}
\usepackage{setspace}

\allowdisplaybreaks

\def\dd{\text{d}}

\begin{document}

\preprint{ZMP-HH/24-5, TCDMATH 24–02}

\title{\Large{A perturbative approach to the non-relativistic string spectrum}}

\author{\bf{Marius de Leeuw}}
\email[E-mail: ]{\tt mdeleeuw[at]maths.tcd.ie} 
\affiliation{\vspace{2mm} School of Mathematics \& Hamilton Mathematics Institute, Trinity College Dublin, Ireland \\ \\
Trinity Quantum Alliance, Unit 16, Trinity Technology and Enterprise Centre, Pearse Street, Dublin 2, Ireland}

\author{\bf{Andrea Fontanella}}
\email[E-mail: ]{\tt andrea.fontanella[at]tcd.ie} 
\affiliation{\vspace{2mm} School of Mathematics $\&$ Hamilton Mathematics Institute, 
Trinity College Dublin, Ireland \\ \\
Perimeter Institute for Theoretical Physics, 
Waterloo, Ontario, N2L 2Y5, Canada}

\author{\bf{Juan Miguel Nieto Garc\'ia}}
\email[E-mail: ]{\tt juan.miguel.nieto.garcia[at]desy.de} 
\affiliation{\vspace{2mm} II. Institut für Theoretische Physik, Universität Hamburg,\\
Luruper Chaussee 149, 22761 Hamburg, Germany}

\begin{abstract}
In this letter we use a perturbative approach to find the spectrum of non-relativistic strings in the String Newton-Cartan (SNC) AdS$_5\times$S$^5$ spacetime. We perturb the bosonic sector of the action around a BMN-like folded string solution in light-cone gauge. We find strong evidence that the theory is described by a combination of massive and massless free fields in an anti-de Sitter background by showing that interaction terms up to six scalars vanish after field redefinitions.
\end{abstract}

\maketitle

\section{Introduction}

A current open problem in theoretical high energy physics is understanding how general is the holographic principle, namely that a theory of gravity is equivalent to a theory of gauge interactions only. The first concrete realisation of holography was proposed by Maldacena \cite{Maldacena:1997re} as a correspondence between type IIB string theory in AdS$_5\times$S$^5$ and $\mathcal{N}=4$ Super Yang-Mills (SYM) theory in 4d flat spacetime. Later, holography was generalised to different spacetime geometries, although an outstanding limitation was that the spacetime always required an Anti de-Sitter (AdS) factor. In recent years, there has been an effort to explore holography in non-AdS spacetimes. Perhaps the most known example is given by flat space holography (for recent reviews on the topic, see e.g. \cite{Raclariu:2021zjz, Donnay:2023mrd, Pasterski:2021raf}).    

Another example of non-AdS holography, perhaps less known, is given by taking the non-relativistic limit of a Lorentzian spacetime. This limit acts on the coordinates of the spacetime and it breaks its Lorentzian structure, therefore rendering it non-AdS.  
In the context of string theory, the non-relativistic limit was first applied to strings propagating in flat spacetime \cite{Gomis:2000bd, Danielsson:2000gi}, and later also in AdS$_5\times$S$^5$ \cite{Gomis:2005pg}. In its original formulation, the non-relativistic limit of a string theory consists in rescaling the embedding coordinates by a dimensionless parameter $c$, together with coupling the string action to a closed critical B-field in order to cancel the usual metric divergence, and finally take $c$ to infinity. In this process the world-sheet remains relativistic, and Weyl anomalies cancel provided the beta-function is set to zero \cite{Gomis:2019zyu,Gallegos:2019icg}. After this procedure, the spacetime geometry changes from being Lorentzian to String Newton-Cartan (SNC) \cite{Andringa:2012uz, Bergshoeff:2018yvt, Bergshoeff:2019pij}. Other approaches to derive the non-relativistic string action are the null reduction \cite{Harmark:2017rpg, Harmark:2018cdl, Harmark:2019upf}, and the expansion method \cite{Hartong:2021ekg, Hartong:2022dsx} (for the application of the latter one to the AdS$_5\times$S$^5$ superstring action, see \cite{Fontanella:2020eje}). For a recent review on aspects of non-relativistic string theory, see \cite{Oling:2022fft}.

Recently, a new holographic correspondence between non-relativistic string and gauge theories was proposed in \cite{Fontanella:2024rvn}. In particular, the proposed duality is between non-relativistic string theory in String Newton-Cartan (SNC) AdS$_5\times$S$^5$ and Galilean Electrodynamics with 5 uncharged massless free scalar fields. It is now an important task to provide a quantitative test of this new non-relativistic duality. One possible test, is the matching of the non-relativistic string spectrum with the conformal dimensions of gauge invariant operators. In the context of the relativistic AdS$_5$/CFT$_4$ correspondence, this program was successfully completed thanks to the fact that theories on both sides of the duality are integrable \cite{Bena:2003wd, Minahan:2002ve}, see \cite{Beisert:2010jr} for a review on the topic. At this point, it is natural to ask whether the theories involved in the non-relativistic holographic correspondence of \cite{Fontanella:2024rvn} are also integrable. There is some evidence pointing out that non-relativistic string theory in SNC AdS$_5\times$S$^5$ is integrable. A non-relativistic version of the Metsaev-Tsytlin action for this theory was constructed in \cite{Fontanella:2022fjd, Fontanella:2022pbm}, which allowed to find the Lax pair \cite{Fontanella:2022fjd}. The spectral curve associated to it has also been studied in \cite{Fontanella:2022wfj}, and it was found to be trivial due to the non semi-simplicity of the non-relativistic algebra. A generalisation of the spectral curve to the non-Lorentzian case was also proposed in \cite{Fontanella:2022wfj}. This fact was already a clear indication that integrability techniques cannot be borrowed directly from the relativistic theory, but they need to be rethought into what it may be called ``non-Lorentzian integrability''.   

The purpose of this letter is to solve perturbatively the spectrum problem of non-relativistic string theory in SNC AdS$_5\times$S$^5$. Our approach consists in fixing light-cone gauge and expanding the action around the BMN-like folded string \cite{Fontanella:2021btt, Fontanella:2023men}. The reasoning of choosing this vacuum is that it is the simplest classical solution to the non-relativistic action that is compatible with the light-cone gauge.\footnote{The relativistic BMN point-like string solution, which is the solution used in the light-cone gauge expansion of the relativistic action, does not survive in the non-relativistic limit \cite{Fontanella:2023men}, and therefore cannot be used for a semiclassical expansion of the non-relativistic action.} In this work, we do not make a particular use of integrability. Instead, the techniques we employ purely consist in field redefinitions and perturbative analysis. Our main result is that the non-relativistic string action in light-one gauge expanded around the BMN-like folded vacuum, after highly non-trivial field redefinitions, becomes the action of free massive and massless scalar fields in AdS$_2$, where all corrections beyond the quadratic order in fields vanish.  Our result holds pertubatively up to, and including, terms of order six in the fields. We will also discuss how our result could be demonstrated at the non-perturbative level for a specific sector of the theory, which requires to solve a complicated PDE.

In the past, the semiclassical expansion of the non-relativistic string action in SNC AdS$_5\times$S$^5$ was already studied in a couple of articles \cite{Fontanella:2021hcb, Sakaguchi:2007ba}. Here we point out the differences between them and the present article. In \cite{Fontanella:2021hcb}, the authors computed the semiclassical expansion of the light-cone gauge fixed action around the ``twisted periodic'' BMN-like folded solution found in \cite{Fontanella:2021btt}. The expansion was done perturbatively in two parameters: a large string tension and also a large radius. In this double expansion, they found the action expands around  free massless scalar fields in Mink$_2$, which is the massless analogue of what happens for the relativistic theory expanded around the BMN vacuum. However, here is where the similarities between the two theories end. The main point of difference is that the relativistic BMN solution is point like, whereas the BMN-like folded solution, and its twisted periodic analogue, are extended objects that necessarily have a spatial longitudinal coordinate that depends non-trivially on $\sigma$ -- the winding. This implies that the expansion of the non-relativistic action around the BMN-like folded solution will have a $\sigma$ dependency, as the spatial longitudinal coordinate is not an isometry. In \cite{Fontanella:2021btt}, the authors found a way to push these $\sigma$ dependent terms to higher orders in perturbation theory. The proposed solution was to expand the action in large string tension, and also in large radius, around the BMN-like folded string with twisted periodic boundary conditions.  
However, even if we push the $\sigma$-dependent terms to higher orders, they still hinder a definition of a perturbative S-matrix, as they become large once the world-sheet is decompactified to define asymptotic scattering states. In this letter, we bypass these issues. We expand the non-relativistic action around the BMN-like folded solution with closed string boundary conditions, and instead of flushing the $\sigma$ dependent terms at higher orders, we make non-trivial field transformations that allow us to recast them into a metric in AdS$_2$. Higher order terms in perturbation theory will not have a $\sigma$-dependence neither, as we show they vanish.

The second article we need to discuss the differences with is \cite{Sakaguchi:2007ba}. In this work, the authors considered the non-relativistic string action in SNC AdS$_5\times$S$^5$, they fixed static gauge and they expanded around the static string solution. 
In this way, they found as a fully non-perturbative result that the expanded action describes free massive and massless fields in AdS$_2$. The main difference between their result and ours is on the choice of vacuum. Their vacuum is the static string solution, which has zero energy and zero linear momentum. Our vacuum is the BMN-like folded string \cite{Fontanella:2021btt}, which has non-trivial energy $E$ and linear momentum $J$ along a flattened out direction of S$^5$, related by a dispersion relation $E \sim J^2$.
Although our result only holds perturbatively in large string tension, the fact we find the correction terms vanish up to the sextic order in the fields is a strong indication that all higher order corrections may vanish as well.  
Then it is rather surprising that the expansion of the non-relativistic action in SNC AdS$_5\times$S$^5$ around the static solution \cite{Sakaguchi:2007ba}, and around the BMN-like folded vacuum (i.e. the present letter), leads to the same spectrum of 3 massive and 5 massless fields in AdS$_2$.

\section{Light-cone gauge}
The theory we consider is the bosonic sector of Type IIB string theory in AdS$_5\times$S$^5$ in the non-relativistic limit. 
We choose to adapt Cartesian coordinates to AdS$_5\times$S$^5$, given by $(t, z_i)$ for AdS$_5$ and $(\phi, y_i)$ for S$^5$,  with $i=1,..., 4$. In these coordinates, the non-relativistic limit is given by rescaling $t$ and one of the $z_i$ coordinates differently from the rest, for concreteness here we take $z_1$. For the full details, we refer the reader to \cite{Fontanella:2021hcb, Fontanella:2021btt}. The procedure of fixing the light-cone gauge for this theory written in Cartesian coordinates has already been shown in \cite{Fontanella:2021hcb}. Here we repeat it briefly for completeness. 

The non-relativistic action is written in the form  
\begin{equation}
\label{NR_action}
S = - \frac{T}{2} \int \dd^2 \sigma \, \bigg( \gamma^{\alpha\beta}\partial_{\alpha} X^{\mu} \partial_{\beta} X^{\nu} H_{\mu\nu} + \lambda_A \mathcal{F}^A \bigg) \ , 
\end{equation}
where
\begin{equation}
 \mathcal{F}^A \equiv \tau_{\mu}{}^A \dot{X}^{\mu} - \frac{1}{\gamma_{11}} \varepsilon^{AB} \eta_{BC} \tau_{\mu}{}^C  X^{\prime \mu} - \frac{\gamma_{01}}{\gamma_{11}} \tau_{\mu}{}^A X^{\prime \mu} \ , 
\end{equation}
and $T$ is the string tension. The string world-sheet coordinates are collectively denoted as $\sigma^{\alpha} = (\tau, \sigma)$, with $\sigma \equiv \sigma + 2 \pi$, $X^{\mu}(\tau, \sigma)$ are the string embedding fields, and $A, B, ...$ are 2d indices taking value $0$ or $1$. The fields $\lambda_A = (\lambda_0,\lambda_1)$ are non-dynamical fields playing the role of Lagrange multipliers, and $\gamma^{\alpha\beta} \equiv \sqrt{-h} h^{\alpha\beta}$ is the Weyl invariant combination of the inverse world-sheet metric $h^{\alpha\beta}$ and $h =$ det$(h_{\alpha\beta})$. Throughout the paper, we use the notation $\dot{X}^{\mu} \equiv \partial_{\tau} X^{\mu}$ and $X^{\prime \mu} \equiv \partial_{\sigma} X^{\mu}$.

The tensors $\tau_{\mu}{}^A$ and $H_{\mu\nu}$ contain the information regarding the target space, which after taking the non-relativistic limit is the SNC AdS$_5\times$S$^5$ manifold. In Cartesian coordinates, where we denote $X^I \equiv (z_i, y_i)$, and after splitting the coordinates $z_i \equiv (z_1, z_m)$, with $m=2,3,4$, these tensors are given by \cite{Fontanella:2021hcb} 
\begin{equation}
H_{\mu\nu} \dd X^{\mu} \dd X^{\nu} =  H_{tt} \dd t^2 + H_{\phi\phi} \dd \phi^2 + H_{IJ} \dd X^I \dd X^J \ , 
\end{equation}
where
\begin{eqnarray}
\notag
H_{tt} &=& - \frac{1+ \frac{z_1^2}{4}}{ \left(1- \frac{z_1^2}{4} \right)^3} z_m z_m \ , \qquad
H_{\phi\phi} = 1 \, , \\
\notag
H_{IJ} \dd X^I \dd X^J &=&  \frac{z_m z_m}{2 \left( 1- \frac{z_1^2}{4} \right)^3} \,  \dd z_1^2 \\
&+& \frac{1}{\left( 1- \frac{z_1^2}{4} \right)^2} \, \dd z_m \dd z_m 
+ \dd y_i \dd y_i \, ,
\end{eqnarray}
and the non-zero components of $\tau_{\mu}{}^A$ are
\begin{equation}
\tau_t{}^0 = -\frac{1+ \frac{z_1^2}{4}}{1- \frac{z_1^2}{4}} \ , \qquad\qquad
\tau_{z_1}{}^1 =\frac{1}{1 - \frac{z_1^2}{4}} \ . \label{tauvielbein}
\end{equation}
In order for us to impose the light-cone gauge, we write the non-relativistic action in the first-order formalism. We introduce the conjugate momenta 
\begin{equation}
\label{def_momenta}
p_{\mu} \equiv \frac{\delta S}{\delta \dot{X}^{\mu}} = - T \gamma^{0\alpha} \partial_{\alpha} X^{\nu} H_{\mu\nu} - \frac{T}{2} \lambda_A \tau_{\mu}{}^A \ .
\end{equation}
The action then takes the form
\begin{equation}
\label{first_NR_action}
S = \int \dd \tau \int_0^{2 \pi} \dd \sigma \, ( p_{\mu} \dot{X}^{\mu} - \mathcal{H} ) \ , 
\end{equation}
where
\begin{eqnarray}
    \mathcal{H} = - \frac{\gamma^{01}}{\gamma^{00}} C_1 - \frac{1}{2 T \gamma^{00}} C_2 \, , 
\end{eqnarray}
is the Hamiltonian, and 
\begin{eqnarray}
C_1 &=& p_{\mu} X^{\prime \mu} \ , \\
\notag
C_2 &=& H^{\mu\nu} p_{\mu}p_{\nu} + T^2 H_{\mu\nu} X^{\prime \mu}X^{\prime\nu} - T^2 \lambda_A \varepsilon^{AB} \eta_{BC} \tau_{\mu}{}^C X^{\prime\mu}  \\
&& + T \lambda_A \tau_{\mu}{}^A p_{\nu} H^{\mu\nu} 
+ \frac{T^2}{4} \lambda_A \lambda_B \tau_{\mu}{}^A \tau_{\nu}{}^B H^{\mu\nu}
 \ ,
\end{eqnarray}
are combinations of the Virasoro constraints. Here $H^{\mu\nu}$ is the usual inverse of $H_{\mu\nu}$, i.e. $H^{\mu\rho}H_{\rho \nu}=\delta^\mu_\nu$.
We refer to \cite{Kluson:2018grx} for its expression in terms of the string Newton-Cartan data.

Next, we introduce the following one-parameter family of light-cone coordinates, 
\begin{eqnarray}
\notag
X_+ &=& (1-a) t + a \phi \ , \qquad\ \ 
X_- = \phi - t \ , \qquad \\
p_+ &=& (1-a) p_{\phi} - a p_t \ , \qquad
p_- = p_{\phi} + p_t \ , 
\end{eqnarray}
where $0 \leq a \leq 1$ parameterises a gauge freedom. We fix light-cone gauge by imposing 
\begin{equation}
\label{lightcone_gauge}
X_+ = \tau \ , \qquad\qquad p_+ = 1 .
\end{equation}
The next step consists in eliminating the Lagrange multiplier fields. Since they are non-dynamical, their conjugate momenta are identically zero. Imposing that this condition is preserved in time, namely 
\begin{equation}
p_{\lambda_A} = 0 \qquad \Longrightarrow \qquad \partial_{\tau} p_{\lambda_A} = \{ p_{\lambda_A}, \mathcal{H}\} \approx 0 \ ,
\end{equation}
will give us two equations, whose solution fixes $\lambda_0$ and $\lambda_1$ in terms of the remaining fields.   
Then, we solve the first Virasoro constraint,
\begin{equation}
C_1 = p_+ X'_- + p_I X^{\prime I}  \approx  0 \ ,  \quad \Longrightarrow \quad X'_- = - p_I X^{\prime I} \ , 
\end{equation}
which fixes $X'_-$ in terms of the remaining fields. Finally, we solve the second Virasoro constraint, i.e. $C_2  \approx  0$, which fixes $p_-$ in terms of the remaining fields. After having eliminated the Lagrange multipliers, $X_-$ and $p_-$, the action (\ref{first_NR_action}) becomes
\begin{eqnarray}
S_{\text{g.f.}} &=& \int \dd^2 \sigma \, (p_I \dot{X}^I - \mathcal{H}_{\text{red}} ) \ ,
\end{eqnarray}
where $\mathcal{H}_{\text{red}} \equiv - p_- (X^I, X^{\prime I}, p_I)$ is the reduced Hamiltonian, which depends on the transverse fields $X^I$ and their momenta only. 

At this stage, where light-cone gauge has been fixed, and all constraints have been solved, we want to eliminate the transverse momenta $p_I$ and obtain an action written in the second order formalism. To do this, we need to compute the Euler-Lagrange equations for the momenta, and solve them for $p_I$ in terms of the transverse fields $X^I$. 
This step, which conceptually is rather simple, is at the same time computationally involved. If $a$ is left generic, this step cannot be done exactly. It can be done perturbatively by expanding around the BMN-like folded vacuum considered in the next section, however, already at the cubic order computations are difficult.   

There is a huge simplification if we consider the gauge\footnote{For general values of $a$, $p_-$ is obtained by solving a quadratic equation. As a consequence of this, eliminating the transverse momenta from the action requires solving a complicated equation involving square roots. In contrast, for $a=0$ the equation for eliminating $p_-$ becomes linear, and there is no square root in the equation for eliminating the transverse momenta. }
\begin{equation}
    a = 0 \, . 
\end{equation}
In this gauge, the momenta can be solved exactly in terms of the transverse fields $X^I$, and the final action is 
\begin{eqnarray}
\label{gauge_fixed_action}
\notag
    S^{a=0}_{\text{g.f.}} &=& \int \dd^2 \sigma \, \frac{1}{8T (z_1^2 - 4)^3 (z_1^2 + 4) z_1^{\prime}} \bigg[ z_1^{10} (1 + T^2 y_j^{\prime 2}) \\
\notag
    &-& 4 z_1^8 (1 + T^2 y_j^{\prime 2} - 2 T z_1^{\prime}) \\
\notag
    &-& 16 z_1^6 \bigg(2 + 4 T z_1^{\prime} - T^2 z_m^{\prime 2} + T^2 z_1^{\prime 2} \dot{y}_j^2 \\
\notag 
    &-& 2 T^2 y_j^{\prime} \dot{y}_j z_1^{\prime}  \dot{z}_1 + \dot{z}_1^2 + T^2 y_j^{\prime 2} (2 + \dot{z}_1^2)\bigg) \\
\notag
    &+& 32 z_1^4 \bigg( 4 - T^2 z_m^2 z_1^{\prime 2} + 2 T^2 z_m^{\prime 2} + 6 T^2 z_1^{\prime 2} \dot{y}_j^2  \\
\notag
    &-& 12 T^2 y_j^{\prime} \dot{y}_j z_1^{\prime}  \dot{z}_1 + 6 \dot{z}_1^2 + 2 T^2 y_j^{\prime 2} (2 + 3 \dot{z}_1^2) \bigg) \\
\notag
    &-& 256 z_1^2 \bigg(-1 + T^2 z_m^{\prime 2} - 6 T^2 y_j^{\prime} \dot{y}_j z_1^{\prime}  \dot{z}_1 + 3 \dot{z}_1^2 \\
\notag
    &+& T^2 y_j^{\prime 2} (-1 + 3 \dot{z}_1^2) - 2 T z_1^{\prime} (2 + T z_m^{\prime} \dot{z}_m \dot{z}_1 ) \\
\notag
    &+& T^2 z_m^{\prime 2} \dot{z}_1^2 + T^2 z_1^{\prime 2} (2 z_m^2 + 3 \dot{y}_j^2 + \dot{z}_m^2 ) \bigg) \\
\notag
    &-& 512 \bigg( 4T^2 y_j^{\prime} \dot{y}_j z_1^{\prime}  \dot{z}_1 - 2 T^2 y_j^{\prime 2} (-1 + \dot{z}_1^2) \\
\notag
    &-& 2(1 + T^2 z_m^{\prime 2})(-1 + \dot{z}_1^2) + 4T z_1^{\prime} (1 + T z_m^{\prime} \dot{z}_m \dot{z}_1 ) \\
    &+& T^2 z_1^{\prime 2} \big(3 z_m^2 - 2 (\dot{y}_j^2 + \dot{z}_m^2) \big) \bigg) \bigg]\, ,  
\end{eqnarray}
where sums over $m$ and $j$ are to be understood.

\section{Expansion around the BMN-like folded vacuum}

The next step is to perturbatively expand the non-relativistic action in light-cone gauge $a=0$ given in (\ref{gauge_fixed_action}). The perturbative parameter is chosen to be the string tension, which we assume to be large. For convenience purposes, we introduce the parameter $\hbar$, related to the string tension by 
\begin{eqnarray}
    \hbar \equiv \frac{1}{\sqrt{T}} \, . 
\end{eqnarray}
For large string tension, we have that  $\hbar$ is small. We should also remember that fixing light-cone gauge implicitly assumes that the vacuum around which we need to expand the action has a unit of linear momentum along the flat direction $\phi$. This means that the static string solution of \cite{Sakaguchi:2007ba} is not suitable for the expansion we are looking for. In the relativistic action, the light-cone gauge is usually associated with choosing the BMN string as the vacuum. However, the BMN string does not have a well-defined non-relativistic limit \cite{Fontanella:2023men}. The right vacuum in this case is given by the BMN-like folded string found in \cite{Fontanella:2021btt}, which is the non-relativistic limit of the non-compact version of the folded string with zero spin \cite{Fontanella:2023men}. The BMN-like folded string that solves the equations of motion of the light-cone gauge action (\ref{gauge_fixed_action}) takes the form   
\begin{eqnarray}
\label{BMNlike_vacuum}
    t= \mu \tau \, , \qquad
    z_1 = 2 \tan \left( -\frac{\kappa}{2} \sigma \right) \, , \qquad
    \phi = \nu \tau \, ,
\end{eqnarray}
with $\kappa \neq 0$ and integer. This solution describes a closed string, as it satisfies the periodic boundary condition, and its energy is proportional to $\nu^2$, i.e., proportional to the square of the linear momentum on the flat direction $\phi$. The solution has 3 parameters $\mu, \kappa, \nu$. However, not all of them are free. We can see that one of the parameters corresponds to a gauge degree of freedom. In particular, imposing the gauge $a=0$, together with the condition $X_+ = \tau$, implies that
\begin{eqnarray}
    a = \frac{\mu - 1}{\mu - \nu} \overset{!}{=} 0  \ , \qquad \Longrightarrow \qquad
    \mu = 1 \, ,
\end{eqnarray}
i.e. we have to fix $\mu = 1$. Moreover, $\nu$ is fixed by the light-cone gauge condition $-\gamma^{00}\dot{\phi}\ T=p_+\overset{!}{=}1$. Only $\kappa \in \mathbb{Z}$ is a free parameter. This matches the number of free parameters of the BMN-like folded string solution in conformal gauge once the angular momentum has been fixed \cite{Fontanella:2021btt}.

Next, to facilitate the computations, it is useful to eliminate the $\hbar$ dependence which appears in several places inside the gauge fixed action (\ref{gauge_fixed_action}). This is achieved in the same way as it was done for the relativistic action \cite{Arutyunov:2009ga}, namely by rescaling the $\sigma$ variable as follows
\begin{eqnarray}
\label{sigma_rescaling}
    \sigma \to \frac{\sigma}{\hbar^2} \, .
\end{eqnarray}
The solution (\ref{BMNlike_vacuum}) continues to be a solution of the equations of motion derived from the action (\ref{gauge_fixed_action}) after implementing the rescaling (\ref{sigma_rescaling}).

Now we are ready to expand the transverse fields $X^I$ in the large string tension ($\hbar \ll 1$) around the BMN-like folded vacuum (\ref{BMNlike_vacuum}),
\begin{eqnarray}\label{exapnsion_BMN_vacuum}
\notag
    z_1 &=& 2 \tan \left( -\frac{\kappa}{2} \sigma \right) + \hbar \, \delta z_1 \ , \\ 
    z_m &=&  \hbar \, \delta z_m \ , \\ 
\notag 
    y_j &=& \hbar \, \delta y_j \ .
\end{eqnarray}
where $\delta z_1, \delta z_m, \delta y_j$ are the fluctuation fields, and they must satisfy periodic boundary conditions. The gauge fixed action will expand in small $\hbar$ as follows
\begin{eqnarray}
\label{action_expansion}
\mathcal{S}^{a=0}_{\text{g.f.}} =  \int \dd^2 \sigma \,  \sum_{j=0}^\infty \hbar^{j-2} \mathcal{L}_j  \, , 
\end{eqnarray}
where $\mathcal{L}_i$ are homogeneous polynomials of degree $i$ in the field fluctuations and their derivatives. In particular, $\mathcal{L}_0$ is a constant, and thus it can be disregarded; $\mathcal{L}_1$ is a total derivative depending linearly on $\delta z_1$ and $\delta z_1^{\prime}$, and gives no contribution to the action thanks to the closed boundary conditions imposed on the field fluctuations. This is consistent with the fact that we are expanding around a solution of the classical equations of motion.

At this stage $\mathcal{L}_2$ appears to be a rather complicated quadratic Lagrangian, depending on $\sigma$ in a cumbersome way. To make progress, we redefine $\tau$ and the fields to take the quadratic Lagrangian to the free fields action in AdS$_2$. The field redefinition is the following,
\begin{align}
\label{redefinition_quadratic}
\notag
    \delta z_1 &\to \sqrt{2} \kappa^{3/2} \sec^2 \left(\frac{\kappa \sigma}{2} \right) \delta z_1 \ , \notag \\
    \delta z_m &\to \frac{\sqrt{2 \kappa}}{\cos^2 (\frac{\kappa \sigma}{2}) \sec (\kappa \sigma)} \delta z_m \ , \\
\notag
    \delta y_j &\to \sqrt{2 \kappa} \, \delta y_j \ , \qquad
    \tau \to \kappa \tau \ ,\notag 
\end{align}
After applying it, the quadratic Lagrangian $\mathcal{L}_2$ describes free fields in AdS$_2$, 
\begin{eqnarray}
\label{quadratic_free_AdS2}
\notag
    &&\mathcal{L}_2 = \sqrt{-g} \bigg[ \frac{1}{2} g^{\alpha\beta}(\partial_{\alpha} \delta z_1 \partial_{\beta} \delta z_1 + \partial_{\alpha} \delta z_m \partial_{\beta} \delta z_m\\  
    && \hspace{2cm}    
    + \partial_{\alpha} \delta y_j \partial_{\beta} \delta y_j) + \kappa^2 \delta z_m^2 \bigg] \ ,
\end{eqnarray}
plus a total derivative which we discarded. Here $g_{\alpha\beta} = \sec (\kappa \sigma)^2 \eta_{\alpha\beta}$ is the AdS$_2$ metric. The quadratic action describes the dynamics of 3 massive fields $\delta z_m$ and 5 massless fields $(\delta z_1, \delta y_j)$.       

Regarding the higher order interaction terms (cubic, quartic, etc.), even after applying the field redefinition (\ref{redefinition_quadratic}), they are still written in a complicated way, depending cumbersomely on $\sigma$. Our guiding principle is to seek a field redefinition that recasts these higher order interactions inside manifestly Lorentz covariant formulas involving the contraction of world-sheet indices with the AdS$_2$ metric $g_{\alpha\beta}$. The general idea is to be able to write down the higher order interaction terms similarly as it was done in \cite{Giombi:2017cqn} for the expansion of the relativistic action around AdS$_2$ minimal surfaces corresponding to boundary Wilson loops.   

The result we obtain is rather surprising. The field redefinition, fixed by demanding covariance and discussed below, shows that all higher interaction terms, up to the sixth order in the fields, vanish.

\subsection{Perturbative field redefinition}

In this section, we shall give the detail of the field redefinition. Inspired by the discussion in Appendix A of \cite{Kruczenski:2004kw}, we transform the field fluctuations perturbatively as follows,
\begin{eqnarray}
\label{trick_full}
\notag
    \delta z_1 &\to& \delta z_1 + \sum_{i=1}^4 \hbar^{i} \, f^{(i)}_{z_1}  \, , \\
    \delta z_m &\to& \delta z_m + \sum_{i=1}^4 \hbar^{i} \, f^{(i)}_{z_m} \, , \\
\notag    
    \delta y_j &\to& \delta y_j + \sum_{i=1}^4 \hbar^{i} \, f^{(i)}_{y_j} \, , 
\end{eqnarray} 
where $f^{(i)}_{z_1}, f^{(i)}_{z_m}, f^{(i)}_{y_j}$ are the coefficients of the field redefinitions at order $\hbar^i$ and are given in Appendix \ref{app:Coefficients}. They depend homogeneously on $i+1$ field fluctuations and have been fixed by demanding Lorentz covariance of the Lagrangian term $\mathcal{L}_{i+2}$. Notice that this field redefinition does not spoil the quadratic Lagrangian (\ref{quadratic_free_AdS2}), which continues to describe free fields in the new field fluctuations. 

The coefficients of the field redefinitions have been computed up to the fourth order, and from their expression we can infer they have the following general pattern:
\begin{eqnarray}
\notag
    \delta z_1 &\to& \delta z_1 + \sum_{n=1}^{\infty}\sum_{i=0}^{n-2} \hbar^{n} A_{i,n} \partial_\sigma^i (\delta z_1^{n+1}) \\
\notag
    && +\sum_{n=1}^{\infty} \hbar^{n} A_{n-1,n} (z_1)_{clas.}\partial_\sigma^{n-2} (\delta z_1^{n} \delta z_1')  \\
    & &+\sum_{n=1}^{\infty} \hbar^{n} B_{n} \partial_\sigma^{n-1} (\delta z_1^{n} \delta z_1') \ , \\
    \delta y_j &\to& \delta y_j + \sum_{n=1}^{\infty} \hbar^{n} B_{n} \partial_\sigma^{n-1} (\delta z_1^{n} \delta y_j') \ , \\
\notag
    \delta z_m &\to& \delta z_m + \sum_{n=1}^{\infty} \sum_{i=0}^{n-1} \sum_{j=0}^{i} \hbar^{n} C_{n,i,j} \partial_\sigma^{i-j} (\delta z_m) \partial_\sigma^j (\delta z_1^{n})\\
    && + \sum_{n=1}^{\infty} \hbar^{n} B_{n} \partial_\sigma^{n-1} (\delta z_1^{n} \delta z_m') \, ,
\end{eqnarray}
where 
\begin{eqnarray}
    (z_1)_{clas.}=2 \tan \left( - \frac{\kappa}{2} \sigma \right) \, ,
\end{eqnarray}
and from our computations up to $\hbar^4$, we know that
\begin{eqnarray}
\notag
    &&B_1=-\sqrt{2\kappa} \, , \quad B_2=\kappa \, , \quad B_3=-\frac{\sqrt{2\kappa^3}}{3} \, , \quad B_4=\frac{\kappa^2}{6} \, , \\
\notag 
     && A_{0,1}=\frac{\sqrt{\kappa^3}}{2\sqrt{2}} \, , \ \, \qquad A_{1,2}=-\kappa^2 \, , \\
     && A_{2,3}=\frac{\sqrt{2\kappa^5}}{8} \, , \qquad A_{3,4}=-\frac{\kappa^3}{15} \ .
\end{eqnarray}
After plugging the field transformation (\ref{trick_full}), with coefficients given in Appendix \ref{app:Coefficients}, into the action (\ref{action_expansion}), we find after some demanding computations that the new higher order interaction terms satisfy 
\begin{eqnarray}
    \mathcal{L}_3 = \mathcal{L}_4 = \mathcal{L}_5 = \mathcal{L}_6 = 0 \, ,  
\end{eqnarray}
up to total derivatives. This result gives strong evidence that the non-relativistic action in light-cone gauge expanded around the BMN-like folded vacuum just describes the dynamics of free fields in AdS$_2$.

\subsection{On the non-perturbative field redefinition}

The result shown previously, namely that we are able to find a perturbative redefinition of the field fluctuations that sets to zero the higher order interactions up to the sextic term, is tantalising that such result might also hold at all orders in $\hbar$. To show that, we would need to find a non-perturbative field redefinition that maps the cumbersome gauge fixed action (\ref{gauge_fixed_action}) to the free fields in AdS$_2$ action. This task is obviously not straightforward, and in this section we detail a systematic approach to this problem, although we are not able to provide the final solution.        

First, we restrict to the sector of the gauge fixed action that contains only $\delta z_1$, by setting $\delta z_m$ and $\delta y_j$ to zero. We introduce a non-perturbative map that replaces $\delta z_1$ by a function $\mathcal{F}$ that, in full generality, depends on the new fluctuation field $\delta z_1$, its world-sheet derivatives and the world-sheet coordinates. Since at the perturbative level the map $\mathcal{F}$ needs to match (\ref{trick_full}), we are suggested to take $\mathcal{F}$ independent of the $\tau$ coordinate, and to disregard any dependence on $\tau$ derivatives of $\delta z_1$. Therefore, our non-perturbative map is of the type
\begin{eqnarray}
\label{non_perturbative}
    \delta z_1 \to \mathcal{F}(\delta z_1, \delta z_1^{\prime}, \delta z_1^{\prime \prime}, ... , \sigma) \, . 
\end{eqnarray}
The idea is that after plugging (\ref{non_perturbative}) inside the gauge fixed action (\ref{gauge_fixed_action}) restricted to the $\delta z_1$ sector, one obtains a theory of massless free field in AdS$_2$.  

Although the gauge fixed action restricted to the $\delta z_1$ sector with the field substitutions (\ref{exapnsion_BMN_vacuum}) and (\ref{redefinition_quadratic}) is still very complicated, we notice that the term $\delta \dot{z}_1^2$ is localised only in one place, with an overall coefficient $K(\delta z_1,\delta z_1^{\prime}, \sigma)$ given by 
\begin{eqnarray}
\label{K_coefficient}
    K(\delta z_1,\delta z_1^{\prime}, \sigma) = \frac{b_1}{b_2 \, b_3} \, ,
\end{eqnarray}
where:
\begin{eqnarray*}
    b_1 &=& - 2 \hbar ^2 \cos ^2\left(\frac{\kappa  \sigma
   }{2}\right) \, , \\
   b_2 &=& 1+ \cos(\kappa \sigma) - \sqrt{2} \kappa^{3/2} \hbar \sin (\kappa \sigma) \delta z_1 + \kappa^3 \hbar^2 \delta z_1^2 \, , \\
   b_3 &=& -1 + \sqrt{2} \kappa^{3/2} \hbar \tan\left(\frac{\kappa  \sigma
   }{2}\right) \delta z_1 + \sqrt{2 \kappa} \hbar \delta z_1^{\prime} \, . 
\end{eqnarray*}
This means that the only place where the new term $\delta \dot{z}_1^2$ can be generated is from $K(\delta z_1,\delta z_1^{\prime}, \sigma) \delta \dot{z}_1^2$. We plug the non-perturbative field transformation into it, and then we demand that the resulting expression gives the free field dynamics, namely we impose
\begin{eqnarray}
\label{condition}
    \int \dd^2 \sigma \, K(\mathcal{F}, \mathcal{F}^{\prime}, \sigma) \dot{\mathcal{F}}^2 =  - \int \dd^2 \sigma \, \delta \dot{z}_1^2 \, . 
\end{eqnarray}
We apply the chain rule to $\dot{\mathcal{F}}^2$ and we integrate by parts all terms of the type $\dot{z}_1 \partial_{\sigma}^{n} \dot{z}_1$, which will contribute to the final $\dot{z}_1^2$ term. In the end, we find that in order for (\ref{condition}) to hold, $\mathcal{F}$ needs to satisfy a PDE with infinite many terms, given by 
\begin{eqnarray}
\label{PDE}
    \sum_{n=0}^{\infty} (-1)^n \partial_{\sigma}^n \left( K(\mathcal{F}, \mathcal{F}^{\prime}, \sigma) \frac{\partial \mathcal{F}}{\partial \delta z_1} \frac{\partial \mathcal{F}}{\partial (\partial_{\sigma}^n \delta z_1)} \right) = - 1 \, . 
\end{eqnarray}
It is an interesting problem to find a solution $\mathcal{F}$ to the PDE (\ref{PDE}), which however we do not plan to pursue here.

\section{Spectrum}\label{sec:Spectrum}

In the previous section, we showed that the fluctuations around the BMN-like folded string are described by free fields in AdS$_2$ up to sixth order in fields. Assuming that this holds at all orders, our task would be to quantise these free fields. This is not as easy as it might appear at first sight, mainly because the AdS space is not globally hyperbolic. Here, we review the results from \cite{PhysRevD.18.3565,SAKAI1985661,Sakaguchi:2007ba} on this topic.

There are two main problems that make the AdS$_n$ space non globally hyperbolic: the presence of closed time-like curves and the existence of a time-like boundary. The first problem can be circumvented by considering the universal covering of the AdS space, where the compact time coordinate is unravelled into the topology of the real line. The second problem is more important and more difficult to solve. The presence of a time-like boundary implies that information can be lost or gained from it. In this situation, the Cauchy problem is ill-defined. One possible way to address this problem is to realise that the universal covering of the AdS space is conformally equivalent to half of the Einstein static universe. Therefore, the solution to the Klein-Gordon equation in the universal covering of AdS differs by a conformal factor from the solution in the Einstein static universe, namely 
\begin{equation}
    \left[ \frac{1}{\sqrt{-\Omega^2 g}}\partial_\mu (\sqrt{-\Omega^2 g} \, \Omega^2 g^{\mu \nu} \partial_\nu ) - \frac{R}{6} \right] \Omega^{-1} \psi (x)=0 \, ,
\end{equation}
where $g_{\mu\nu}$ is the metric of either the universal covering of AdS or the Einstein static universe, $R$ is its curvature scalar and $\Omega$ is the conformal factor taking one geometry to the other one. We can use this fact to solve the equations of motion associated to (\ref{quadratic_free_AdS2}) by solving instead the equations of motion in the Einstein static universe, where the problem is well posed because the space is globally hyperbolic, imposing that there is no probability flux though the boundary, and undoing the conformal transformation. Although we cannot conformally map massive free fields in AdS to massive free fields in the Einstein static universe, we can impose the same condition at the boundary.

After applying the above argument, we find that the field operator $\psi$ that fulfils the differential equation
\begin{equation}
    (-\partial_\tau^2 + \partial_\sigma^2) \psi = \frac{m^2 \kappa^2}{\cos ^2 \kappa \sigma} \psi \, , \label{EoMfield}
\end{equation}
takes the form
\begin{align}
    \psi^\pm=\sum_{n=0}^\infty N_n &\left[ e^{i (n+\Delta^\pm) \kappa \tau} (\cos \kappa \sigma )^{\Delta^\pm} C^{\Delta^\pm}_n (\sin \kappa \sigma)\, \hat{a}_n^{\phantom{\dagger}}  \right. \notag\\
    +& \left. e^{-i (n+\Delta^\pm) \kappa \tau} (\cos \kappa \sigma )^{\Delta^\pm} C^{\Delta^\pm}_n (\sin \kappa \sigma)\, \hat{a}_n^\dagger \right] \, , \label{field}
\end{align}
where $C^p_q(x)$ are the Gegenbauer polynomials, $\hat{a}_n $ and $\hat{a}_n^\dagger$ are operators that satisfy canonical bosonic commutation relations, and
\begin{equation}
    N_n = \frac{ \Gamma(\Delta) 2^{\Delta -1}}{\sqrt{\pi}} \sqrt{\frac{n!}{\Gamma (n+2\Delta)}} \, , \quad \Delta^{\pm}= \frac{1 \pm \sqrt{1+4 m^2}}{2} \, .
\end{equation}
This solution is normalizable only if the corresponding $\Delta$ is real and larger than $-1/2$, which is equivalent to the Breitenlohner-Freedman bound $m^2 \geq -1/4$ \cite{BREITENLOHNER1982197,BREITENLOHNER1982249}. At this point, the construction of the Fock space proceeds as in flat space.

The equations of motion for the fluctuations in the Lagrangian (\ref{quadratic_free_AdS2}) are all of the form (\ref{EoMfield}) with either $m^2=0$ or $m^2=2$. Thus, the Fock space of our system is constructed as follows. We define first the vacuum state $|0\rangle$ as the state that is annihilated by all the annihilation operators
\begin{equation}
    \hat{a}_n |0\rangle =\hat{b}_{m,n} |0\rangle =\hat{c}_{j,n} |0\rangle =0 \qquad \forall n\in \mathbb{N}_0 \, ,
\end{equation}
where the oscillators $a$, $b$ and $c$ correspond to the fluctuations $\delta z_1$, $\delta z_m$ and $\delta y_i$ respectively. Notice that $a$ and $c$ are excitations with $m^2=0$ while $b$ has $m^2=2$. We can now define the Fock space as
\begin{equation}
    \mathcal{F}=\bigoplus_{p_0, p_1 \dots}  \prod_{n=0}^\infty \left[ (\hat{a}^\dagger_n)^{p_n} \left(\prod_m (\hat{b}^\dagger_{m,n})^{q_n} \right) \left(\prod_j (\hat{c}^\dagger_{j,n})^{r_n} \right) \right] |0\rangle \ .
\end{equation}
Finally, we derive the Hamiltonian associated to (\ref{quadratic_free_AdS2}), we substitute the field operator (\ref{field}), and we get 
\begin{align}
    H=\sum_{n=0}^\infty &\left[ (n+1) \left(\hat{a}_{n}^\dagger \hat{a}_{n}^{\phantom{\dagger}} +\sum_j \hat{c}_{j,n}^\dagger \hat{c}_{j,n}^{\phantom{\dagger}} \right) \right.  \notag \\
    &\left.+\left(  n+2 \right) \sum_m \hat{b}_{m,n}^\dagger \hat{b}_{m,n}^{\phantom{\dagger}} \right] \ ,
\end{align}
up to a zero-point energy, which should cancel once supersymmetry is implemented. Thus, the fluctuations arrange themselves into discrete energy levels given by 
\begin{eqnarray}
    E_n = n+\Delta^+ \, . 
\end{eqnarray}

\section{Conclusions}

In this letter we have studied the light-cone gauge quantisation of non-relativistic string theory in the SNC AdS$_5\times$S$^5$ background expanded around the BMN-like folded string of \cite{Fontanella:2021btt}. We have found that, at large tension and in the decompactification limit, the gauge-fixed Lagrangian is equivalent to the Lagrangian of free fields in AdS$_2$ up to and including the fourth-order in perturbation theory (i.e., the sextic interaction term) after a perturbative redefinition of fields. If this holds at all orders, this would imply that the fluctuations around the BMN-like folded string have a discrete energy spectrum, which we discussed in Section \ref{sec:Spectrum}.

It is surprising that the spectrum we find here is exactly the same as the spectrum of excitations of (\ref{NR_action}) in static gauge expanded around the static solution, which was computed in \cite{Sakaguchi:2007ba}. Although the results are the same, we are computing fluctuations around different vacua, so there is no a priori reason for this matching. It would be interesting to study fluctuations around other classical solutions to see if this is an accident or a universal property, similarly to relativistic strings in flat spacetime where the action is Gaussian (and therefore the spectrum of fluctuations around any classical string solution is the same). 

It would be interesting to prove our result to all orders by finding a non-perturbative field redefinition. We expect this to be possible, because strings in SNC AdS$_5\times$S$^5$ admit a Lax representation \cite{Fontanella:2022fjd} which indicates that they are classically integrable. In the absence of a quartic term in the action, it seems impossible to write down an action with higher order interaction terms that does not allow for particle production already at tree-level \cite{Dorey:1996gd}. The fact that the sextic order vanishes too is consistent with this interpretation.

The analysis of this letter only involves the bosonic sector of the non-relativistic Type IIB superstring theory. It would be a natural next step to include fermions. However, since we have indications that supersymmetry is not broken \cite{Gomis:2005pg}, we expect fermions to contribute to the quadratic order at most.

Given the connection between the BMN-like folded vacuum and the non-compact folded string found in \cite{Fontanella:2023men}, it would be interesting to study relativistic strings in AdS$_5\times$S$^5$ expanded around the non-compact folded string. In light of the results of this paper, it might be possible to obtain huge simplifications in that context and give new insights into the relativistic theory. It would also be interesting to explore if the non-relativistic limit can directly be taken on the level of physical quantities such as the spectrum.

As pointed out in \cite{Gomis:2005pg}, the spectrum of non-relativistic strings in SNC AdS$_5\times$S$^5$ is T-dual to the spectrum obtained from the discrete light-cone quantisation (DLCQ) of relativistic Type IIA string theory in a time dependent pp-wave background. Therefore, the difficulties of such computation, see e.g. \cite{Komatsu:2024vnb} in a related context, can be bypassed by borrowing the result of this letter.

Finally, it would be important to understand the field theoretic counterpart of the results of this letter in view of the recently proposed holographic duality \cite{Fontanella:2024rvn}.

\section{\label{sec:ackn}Acknowledgments}

We thank S. Frolov for useful discussions. MdL was supported in part by SFI and the Royal Society for funding under grants UF160578, RGF$\backslash$ R1$\backslash$ 181011, RGF$\backslash$ EA$\backslash$ 180167 and RF$\backslash$ ERE$\backslash$ 210373. MdL is also supported by ERC-2022-CoG - FAIM 101088193. MdL thanks the Perimeter Institute for hosting him during the beginning of the project.  
During the first part of this work, AF was supported by Perimeter Institute. Research at Perimeter Institute is supported in part by the Government of Canada through the Department of Innovation, Science and Economic Development and by the Province of Ontario through the Ministry of Colleges and Universities. During the last part of this work, AF was supported by SFI and the Royal Society under the grant number RFF$\backslash$EREF$\backslash$210373. JMNG is supported by the Deutsche Forschungsgemeinschaft (DFG, German Research Foundation) under Germany's Excellence Strategy -- EXC 2121 ``Quantum Universe'' -- 390833306.
AF thanks Lia for her permanent support.


\bibliographystyle{nb}

\bibliography{Biblio.bib}


\onecolumngrid
\appendix

\section{Conventions}
\label{app:Conventions}

For a generic object $\mathcal{O}^A$, we define its light-cone combinations as
\begin{equation}\label{LC_comb}
\mathcal{O}^{\pm} \equiv \mathcal{O}^0 \pm \mathcal{O}^1 \ , \qquad\qquad
\mathcal{O}_{\pm} \equiv \frac{1}{2}\left( \mathcal{O}_0 \pm \mathcal{O}_1 \right) \ .
\end{equation}
The 2-dimensional Minkowski metric in light-cone coordinates has non-vanishing components $\eta_{+-} = -1/2$ and $\eta^{+-} = -2$. Our convention for the Levi-Civita tensor is $\varepsilon^{01} = - \varepsilon_{01} = + 1$, which in light-cone coordinates reads $\varepsilon_{+-} = \frac{1}{2}$, $\varepsilon^{+-} = -2$.
Our convention for $p$-forms is $\omega_p = \frac{1}{p!} \omega_{\mu_1 \cdots \mu_p} \dd x^{\mu_1}\wedge \cdots \wedge \dd x^{\mu_p}$.

\section{Coefficients of the perturbative field redefinition}
\label{app:Coefficients}

The coefficients entering the perturbative field redefinition (\ref{trick_full}) are:  

\begin{itemize}

\item  $\mathcal{O}(\hbar)$ 
\begin{eqnarray}
  f^{(1)}_{z_1} &=& - \frac{\kappa^{3/2}}{\sqrt{2}} \tan \left( \frac{\kappa \sigma}{2}\right) \delta z_1^2 - \sqrt{2 \kappa} \, \delta z_1 \delta z_1' \ , \\
  f^{(1)}_{z_m} &=& \sqrt{2} \kappa^{3/2}\sec \left(\kappa \sigma\right)\tan \left( \frac{\kappa \sigma}{2}\right) \delta z_1 \delta z_m - \sqrt{2 \kappa} \, \delta z_1 \delta z_m' \ , \\
  f^{(1)}_{y_j} &=& - \sqrt{2 \kappa} \, \delta z_1 \delta y_j' \ .
\end{eqnarray}

\item  $\mathcal{O}(\hbar^2)$
\begin{eqnarray}
\notag
    f^{(2)}_{z_1} &=& \left[-\frac{\kappa^3}{3} + \frac{ \kappa^3}{2} \sec^2 \left(\frac{\kappa \sigma}{2}\right) \right] \delta z_1^3 + 2 \kappa^2 \tan \left(\frac{\kappa \sigma}{2}\right)\delta z_1^2 \delta z_1^{\prime} \\
    & &+2 \kappa \delta z_1 \left(\delta z_1^{\prime}\right)^2 + \kappa \delta z_1^2 \delta z_1^{\prime \prime} \, ,  \\[0.4cm]
    \notag
    f^{(2)}_{z_m} &=& \frac{\kappa^3 \left(1 - 2\sec (\kappa \sigma) \right)}{1 + \cos(\kappa \sigma)} \delta z_1^2 \delta z_m -2\kappa^2 \sec \left(\kappa \sigma\right) \tan \left(\frac{\kappa \sigma}{2}\right) \delta z_1^2 \delta z_m^{\prime} \\
    & &-2\kappa^2 \sec \left(\kappa \sigma\right) \tan \left(\frac{\kappa \sigma}{2}\right) \delta z_1 \delta z_1^{\prime} \delta z_m + 2 \kappa \delta z_1 \delta z_1^{\prime} \delta z_m^{\prime} + \kappa \delta z_1^2 \delta z_m^{\prime \prime} \, , \\[0.4cm]
    f^{(2)}_{y_j} &=& 2 \kappa \delta z_1 \delta z_1^{\prime} \delta y_j^{\prime} + \kappa \delta z_1^2 \delta y_j^{\prime \prime} \, .
\end{eqnarray}

\item  $\mathcal{O}(\hbar^3)$ 
\begin{eqnarray}
\notag
    f^{(3)}_{z_1} &=& -2 \sqrt{2} \kappa^{3/2} (\delta z_1^{\prime})^3 \delta z_1 - 3 \sqrt{2}  \kappa^{3/2} \delta z_1^{\prime \prime} \delta z_1^{\prime} \delta z_1^2 - \frac{\sqrt{2}}{3}  \kappa^{3/2} \delta z_1^{\prime \prime \prime}\delta z_1^3 \\
    \notag
    & & - 3 \sqrt{2} \kappa^{5/2} \tan \left(\frac{\kappa \sigma}{2}\right) (\delta z_1^{\prime})^2 \delta z_1^2 - \sqrt{2} \kappa^{5/2} \tan \left(\frac{\kappa \sigma}{2}\right) \delta z_1^{\prime \prime} \delta z_1^3 \\
    \notag
    & &  + \frac{\kappa^{9/2}}{24 \sqrt{2}} \sec^3 \left(\frac{\kappa \sigma}{2}\right) \left[-11 \sin \left(\frac{\kappa \sigma}{2}\right) + \sin \left(\frac{3 \kappa \sigma}{2}\right) \right] \delta z_1^4  \\
    & & + \sqrt{2} \kappa^{7/2} \frac{\cos (\kappa \sigma) - 2}{\cos (\kappa \sigma) + 1} \delta z_1^{\prime} \delta z_1^3 \, , \\[0.4cm]
\notag
    f^{(3)}_{z_m} &=& - \frac{\sqrt{2}}{3}  \kappa^{3/2} \delta z_m^{\prime \prime \prime}\delta z_1^3 - \sqrt{2} \kappa^{3/2}\delta z_m^{\prime} \delta z_1^{\prime\prime} \delta z_1^2 - 2\sqrt{2} \kappa^{3/2} \delta z_m^{\prime \prime}  \delta z_1^{\prime}\delta z_1^2 \\
    \notag 
    & & - 2\sqrt{2} \kappa^{3/2} \delta z_m^{\prime} (\delta z_1^{\prime})^2 \delta z_1 + \sqrt{2} \kappa^{5/2} \sec (\kappa \sigma) \tan \left(\frac{\kappa \sigma}{2}\right) \delta z_m^{\prime \prime} \delta z_1^3 \\
    \notag 
    & & + 4 \sqrt{2} \kappa^{5/2} \sec (\kappa \sigma) \tan \left(\frac{\kappa \sigma}{2}\right) \delta z_m^{\prime} \delta z_1^{\prime} \delta z_1^2
    \\
    \notag 
    & & + 2\sqrt{2} \kappa^{5/2} \sec (\kappa \sigma) \tan \left(\frac{\kappa \sigma}{2}\right) \delta z_m (\delta z_1^{\prime})^2 \delta z_1  \\
    \notag 
    & & + \sqrt{2} \kappa^{5/2} \sec (\kappa \sigma) \tan \left(\frac{\kappa \sigma}{2}\right) \delta z_m \delta z_1^{\prime \prime} \delta z_1^2 \\
    \notag
    & & + \frac{\kappa^{7/2}}{\sqrt{2}} \sec^2 \left(\frac{\kappa \sigma}{2}\right) \left( 2\sec (\kappa \sigma) - 1 \right)  \delta z_m^{\prime} \delta z_1^3 \\
    \notag
    & & + \sqrt{2} \kappa^{7/2} \sec^2 \left(\frac{\kappa \sigma}{2}\right) \left( 2\sec (\kappa \sigma) - 1 \right) \delta z_m \delta z_1^{\prime} \delta z_1^2 \\
    & & - \frac{\kappa^{9/2}}{6 \sqrt{2}} \sec^3 \left(\frac{\kappa \sigma}{2}\right) \sec (\kappa \sigma) \left[-11 \sin \left(\frac{\kappa \sigma}{2}\right) + \sin \left(\frac{3\kappa \sigma}{2}\right) \right] \delta z_m  \delta z_1^3 \, , \qquad\\[0.4cm]
\notag 
     f^{(3)}_{y_j} &=& - \frac{\sqrt{2}}{3} \kappa^{3/2} \delta z_1^3 \delta y_j^{\prime \prime \prime} - \sqrt{2} \kappa^{3/2} \delta z_1^2  \delta z_1^{\prime \prime} \delta y_j^{\prime} \\
     & & - 2 \sqrt{2} \kappa^{3/2} \delta z_1^2 \delta z_1^{\prime}\delta y_j^{\prime \prime} - 2 \sqrt{2} \kappa^{3/2} \delta z_1 (\delta z_1^{\prime})^2 \delta y_j^{\prime} \, . 
\end{eqnarray}
\item  $\mathcal{O}(\hbar^4)$
\begin{eqnarray}
\notag
    f^{(4)}_{z_1} &=& 4 \kappa^2 (\delta z_1^{\prime})^4 \delta z_1 + 12 \kappa^2 (\delta z_1^{\prime})^2 \delta z_1^{\prime \prime} \delta z_1^2 + 2 \kappa^2 (\delta z_1^{\prime \prime})^2 \delta z_1^3 + \frac{8}{3} \kappa^2 \delta z_1^{\prime \prime \prime} \delta z_1^{\prime} \delta z_1^3 \\
    \notag 
    & & +\frac{1}{6} \kappa^2 \delta z_1^{\prime \prime \prime \prime} \delta z_1^4 + 8 \kappa^3 \tan \left(\frac{\kappa \sigma}{2}\right) (\delta z_1^{\prime})^3 \delta z_1^2 + 8 \kappa^3 \tan \left(\frac{\kappa \sigma}{2}\right) \delta z_1^{\prime \prime} \delta z_1^{\prime} \delta z_1^3  \\
    \notag 
    & & +  \frac{2}{3} \kappa^3 \tan \left(\frac{\kappa \sigma}{2}\right)\delta z_1^{\prime \prime \prime} \delta z_1^4 - 2 \kappa^4 (\cos(\kappa \sigma) - 2) \sec^2 \left(\frac{\kappa \sigma}{2}\right) (\delta z_1^{\prime})^2 \delta z_1^3 \\
    \notag 
    & & +\kappa^4 \frac{2 -\cos(\kappa \sigma)}{1 + \cos(\kappa \sigma)} \delta z_1^{\prime \prime}\delta z_1^4 \\
    \notag
    & & - \frac{\kappa^5}{6} \sec^3 \left(\frac{\kappa \sigma}{2}\right) \left[ -11 \sin \left(\frac{\kappa \sigma}{2}\right) + \sin \left(\frac{3 \kappa \sigma}{2}\right) \right] \delta z_1^{\prime} \delta z_1^4 \\
    & & + \frac{\kappa^6}{240} \left[33 - 26 \cos (\kappa \sigma) + \cos (2\kappa \sigma) \right] \sec^4 \left(\frac{\kappa \sigma}{2}\right) \delta z_1^5 \, , \\[0.4cm]
\notag 
    f^{(4)}_{z_m} &=& -\frac{\kappa^6}{48} \left[ 33 - 26 \cos(\kappa \sigma) + \cos(2\kappa \sigma) \right] \sec^4 \left(\frac{\kappa \sigma}{2}\right) \sec (\kappa \sigma) \delta z_1^4 \delta z_m \\
    \notag 
    & & + \frac{\kappa^5}{6} \sec^3 \left(\frac{\kappa \sigma}{2}\right)\sec (\kappa \sigma) \left[ -11 \sin \left(\frac{\kappa \sigma}{2}\right) + \sin \left(\frac{3 \kappa \sigma}{2}\right) \right] \delta z_1^4 \delta z_m^{\prime} \\
    \notag 
    & & + \frac{\kappa^5}{2} \sec^3 \left(\frac{\kappa \sigma}{2}\right)\sec (\kappa \sigma) \left[ -11 \sin \left(\frac{\kappa \sigma}{2}\right) + \sin \left(\frac{3 \kappa \sigma}{2}\right) \right] \delta z_1^3 \delta z_1^{\prime} \delta z_m \\
    \notag 
    & & + \kappa^4 \frac{1 - 2\sec(\kappa \sigma)}{1 + \cos (\kappa \sigma)} \delta z_1^4 \delta z_m^{\prime \prime} + 3 \kappa^4 \sec^2 \left(\frac{\kappa \sigma}{2}\right) \left[ 1 -2\sec(\kappa \sigma)\right] \delta z_1^3 \delta z_1^{\prime} \delta z_m^{\prime} \\
    \notag 
    & & + \kappa^4 \sec^2 \left(\frac{\kappa \sigma}{2}\right) \left[ 1 -2\sec(\kappa \sigma)\right] \delta z_1^3 \delta z_1^{\prime\prime} \delta z_m \\
    \notag
    & & + 3 \kappa^4 \sec^2 \left(\frac{\kappa \sigma}{2}\right) \left[ 1 -2\sec(\kappa \sigma)\right] \delta z_1^2 (\delta z_1^{\prime})^2 \delta z_m \\
    \notag 
    & & - \frac{2 \kappa^3}{3} \sec(\kappa \sigma) \tan \left(\frac{\kappa \sigma}{2}\right) \delta z_1^4 \delta z_m^{\prime\prime\prime}
    + 6 \kappa^3 \sec(\kappa \sigma) \tan \left(\frac{\kappa \sigma}{2}\right) \delta z_1^3 \delta z_1^{\prime} \delta z_m^{\prime\prime} \\
    \notag 
    & & + 4 \kappa^3 \sec(\kappa \sigma) \tan \left(\frac{\kappa \sigma}{2}\right) \delta z_1^3 \delta z_1^{\prime\prime} \delta z_m^{\prime} + 12 \kappa^3 \sec(\kappa \sigma) \tan \left(\frac{\kappa \sigma}{2}\right) \delta z_1^2 (\delta z_1^{\prime})^2 \delta z_m^{\prime} \\
    \notag 
    & & - \frac{2 \kappa^3}{3} \sec(\kappa \sigma) \tan \left(\frac{\kappa \sigma}{2}\right) \delta z_1^3 \delta z_1^{\prime\prime\prime} \delta z_m 
    - 6 \kappa^3 \sec(\kappa \sigma) \tan \left(\frac{\kappa \sigma}{2}\right) \delta z_1^2 \delta z_1^{\prime} z_1^{\prime\prime} \delta z_m \\
    \notag 
    & & - 4 \kappa^3 \sec(\kappa \sigma) \tan \left(\frac{\kappa \sigma}{2}\right) \delta z_1 (\delta z_1^{\prime})^3  \delta z_m + \frac{\kappa^2}{6} \delta z_1^4 \delta z_m^{\prime \prime\prime\prime} + 2 \kappa^2 \delta z_1^3 \delta z_1^{\prime} \delta z_m^{\prime \prime\prime} \\
    \notag 
    & & + 2 \kappa^2 \delta z_1^3 \delta z_1^{\prime\prime} \delta z_m^{\prime \prime} + 6 \kappa^2 \delta z_1^2 (\delta z_1^{\prime})^2 \delta z_m^{\prime \prime} + 4 \kappa^2 \delta z_1 (\delta z_1^{\prime})^3 \delta z_m^{\prime} \\
    & & + 6 \kappa^2 \delta z_1^2 \delta z_1^{\prime\prime} \delta z_1^{\prime} \delta z_m^{\prime} + \frac{2\kappa^2}{3} \delta z_1^3 \delta z_1^{\prime\prime\prime} \delta z_m^{\prime} \, , \\[0.4cm]
\notag 
    f^{(4)}_{y_j} &=& \frac{\kappa^2}{6} \delta z_1^4 \delta y_j^{\prime\prime\prime\prime} + 2 \kappa^2 \delta z_1^3  \delta z_1^{\prime} \delta y_j^{\prime\prime\prime} + 2\kappa^2 \delta z_1^3  \delta z_1^{\prime\prime} \delta y_j^{\prime\prime} + 6\kappa^2 \delta z_1^2  (\delta z_1^{\prime})^2 \delta y_j^{\prime\prime} \\
    & & + 4\kappa^2 \delta z_1  (\delta z_1^{\prime})^3 \delta y_j^{\prime} + 6\kappa^2 \delta z_1^2  \delta z_1^{\prime}\delta z_1^{\prime\prime} \delta y_j^{\prime} + \frac{2 \kappa^2}{3} \delta z_1^3  \delta z_1^{\prime\prime \prime} \delta y_j^{\prime} \, .  
\end{eqnarray}
    
\end{itemize}

\end{document}